\documentclass[preprint2]{aastex}

\usepackage{bm}

\def\la{\;
\raise0.3ex\hbox{$<$\kern-0.75em\raise-1.1ex\hbox{$\sim$}}\; }
\def\ga{\;
\raise0.3ex\hbox{$>$\kern-0.75em\raise-1.1ex\hbox{$\sim$}}\; }

\newcommand{\hhho}{{H$_3$O$^+$}}
\newcommand{\nhhh}{{NH$_3$}}
\newcommand{\etal}{{et al.}}
\newcommand{\fref}[1]{Fig.~\ref{#1}}

\newcommand{\Eref}[1]{Eq.~(\ref{#1})}
\newcommand{\tref}[1]{Table~\ref{#1}}

\newcommand{\cm}{cm$^{-1}$}
\newcommand{\ms}{m~s$^{-1}$}
\newcommand{\kms}{km~s$^{-1}$}

\slugcomment{Submitted to ApJ}

\shorttitle{$m_{\rm e}/m_{\rm p}$ constraints}
\shortauthors{Kozlov \& Levshakov}

\begin{document}


\title{Sensitivity of the H$_3$O$^+$ inversion-rotational spectrum to changes
in the electron-to-proton mass ratio}


\author{M. G. Kozlov}
\affil{Petersburg Nuclear Physics Institute, Gatchina 188300, Russia}
\email{mgk@mf1309.spb.edu}

\and

\author{S. A. Levshakov\altaffilmark{1}}
\affil{Key Laboratory for Research in Galaxies and Cosmology,
Shanghai Astronomical Observatory, Shanghai 200030, P.R. China}
\email{lev@astro.ioffe.rssi.ru}


\altaffiltext{1}{On leave from A. F. Ioffe Physical-Technical Institute, Saint Petersburg 194021, Russia}


\begin{abstract}
Quantum mechanical
tunneling inversion transition in ammonia \nhhh\ is actively used as a sensitive
tool to study possible variations of the electron-to-proton mass ratio,
$\mu = m_{\rm e}/m_{\rm p}$.
The molecule \hhho\ has the inversion barrier significantly
lower than that of \nhhh. Consequently, its tunneling transition
occurs in the far-infrared (FIR) region and mixes with rotational
transitions. Several such FIR and submillimiter transitions are
observed from the interstellar medium in the Milky Way and in nearby
galaxies. We show that the rest-frame frequencies of these
transitions are very sensitive to the variation of $\mu$,
and that
their sensitivity coefficients have \textit{different} signs. Thus,
\hhho\ can be used as an independent target to test hypothetical
changes in $\mu$ measured at different ambient conditions of high
(terrestrial) and low (interstellar medium) matter densities. The
environmental dependence of $\mu$ and coupling constants is
suggested in a class of chameleon-type scalar field models~---
candidates to dark energy carrier.
\end{abstract}


\section{Introduction}
\label{sect-1}

The spatial and temporal variability of dimensionless physical
constants has become a topic of considerable interest in laboratory
and astrophysical studies as a test of the Einstein equivalence
principle of local position invariance (LPI), which states that
outcomes of nongravitational experiments should be independent of
their position in space-time (e.g., Dent 2008).
The violation of LPI is anticipated in some extensions of Standard Model
and, in particular, in those dealing with dark energy 
(e.g., Hui \etal\ 2009; Damour \& Donoghue 2010).

A concept of dark energy with negative pressure ($p = -\rho$)
appeared in physics long before the discovery of the accelerating
universe through observations of nearby and distant (at redshif $z
\sim 1$) supernovae type Ia \citep{Per98, Rie98}. 
Examples of dark energy in a form of a scalar field with a self-interaction
potential can be found in reviews by \cite{PR03}, \cite{Cop06}, and by \cite{Uz10}. 
Since that time many sophisticated models have been suggested to explain the nature of
dark energy and among them the scalar fields which are ultra-light
in cosmic vacuum but possess a large mass locally when they are
coupled to ordinary matter by the so-called chameleon mechanism
\citep{KW04, Br04, Ave08, Br10a, Br10b}. 
A subclass of these models
considered by \cite{OP08} predicts that fundamental physical
quantities such as elementary particle masses and low-energy
coupling constants may also depend on the local matter density.
Since the mass of the electron $m_{\rm e}$ is proportional to the
Higgs vacuum expectation value (VEV $\sim 200$ GeV), and the mass of
the proton $m_{\rm p}$ is proportional to the quantum chromodynamics
(QCD) scale $\Lambda_{\rm QCD} \sim 220$ MeV, we may probe the ratio
of the electroweak scale to the strong scale through the
measurements of the dimensionless mass ratio $\mu = m_{\rm e}/m_{\rm p}$ 
in high density laboratory (terrestrial) environment, $\mu_{\rm lab}$, 
and in low density interstellar clouds, $\mu_{\rm space}$
($\rho_{\rm lab}/\rho_{\rm space} > 10^{10}$). In this way we are
testing whether the scalar field models have chameleon-type
interaction with ordinary matter.
Several possibilities to detect chameleons from astronomical observations
were discussed in \cite{BDS09}, \cite{DSS09}, \cite{BZ10}, and \cite{Av10}.
First experiments constraining these models were recently carried out 
in Fermilab \citep{USW10} and in Lawrence Livermore National Laboratory \citep{Ryb10}.  

At the moment, the most accurate relative changes in the mass ratio
$\Delta \mu/\mu = (\mu_{\rm space} - \mu_{\rm lab})/\mu_{\rm lab}$
can be obtained with the ammonia method \citep{VKB04,FK07}. 
\nhhh\ is a molecule whose inversion frequencies are very sensitive to any
changes in $\mu$ because of the quantum mechanical tunneling of the
N atom through the plane of the H atoms. The sensitivity coefficient
to $\mu$-variation of the \nhhh\ $(J,K) = (1,1)$ inversion
transition at 24 GHz is $Q_{\rm inv}=4.46$. This means that
the inversion frequency scales as $\Delta \omega/\omega = 4.46(\Delta
\mu/\mu)$. In other words, sensitivity to $\mu$-variation is 4.46
times higher than that of molecular rotational
transitions, where $Q_{\rm rot}=1$. Thus, by comparing the observed
radial velocity of the inversion transition of \nhhh, $V_{\rm inv}$,
with a suitable rotational transition, $V_{\rm rot}$, of another
molecule arising co-spatially with ammonia, a limit on the spatial
variation of $\mu$ can be determined:
\begin{equation}\label{eq1}
 \frac{\Delta \mu}{\mu}
 = \frac{V_{\rm rot} - V_{\rm inv}}{c(Q_{\rm inv} - Q_{\rm rot})}
 \approx 0.3 \frac{\Delta V}{c},
 \end{equation}
where $c$ is the speed of light and $\Delta V = V_{\rm rot} - V_{\rm inv}$.

Surprisingly, recent observations of a sample of nearby (distance $R
\sim$ 140 pc) cold molecular cores ($T_{\rm kin} \sim 10$K, $n =
10^4-10^5$ cm$^{-3}$, $B < 10$ $\mu$G) in lines of \nhhh\ $(J,K) =
(1,1)$ at 24 GHz, HC$_3$N $J = 2-1$ at 18 GHz, and N$_2$H$^+$ $J =
1-0$ at 93 GHz reveal a statistically significant positive velocity
offset between the low-$J$ rotational and inversion transitions:
$\Delta V = V_{\rm rot} - V_{\rm inv} = 27 \pm 4_{\rm stat} \pm
3_{\rm sys}$ m~s$^{-1}$, which gives $\Delta \mu/\mu = (26 \pm
4_{\rm stat} \pm 3_{\rm sys})\times 10^{-9}$ \citep{LML10}\footnote{Presented are the
corrected values of $\Delta V$ and $\Delta \mu/\mu$ discussed in \cite{LLH10}.}. 
A few molecular cores from this sample were mapped in the \nhhh\ (1,1) and
HC$_3$N (2--1) lines and it was found that in two of them (L1498 and
L1512) these lines trace the same material and show the offset
$\Delta V = 26.9 \pm 1.2_{\rm stat} \pm 3.0_{\rm sys}$  m~s$^{-1}$
throughout the entire clouds \citep{LLH10}. It was also demonstrated
that for these clouds the frequency shifts caused by external
electric and magnetic fields and by the cosmic black body
radiation-induced Stark effect are less than 1 \ms.
Optical depth effects in these clouds were studied from the
analysis of unsaturated ($\tau < 1$) and slightly saturated ($\tau \approx 1-2$)
hyperfine components of the corresponding molecular transitions and
it was found that both groups of lines have similar radial
velocities within the $1\sigma$ uncertainty intervals.

The nonzero $\Delta\mu$ implies that at deep interstellar vacuum the
electron-to-proton mass ratio increases by $\sim 3\times10^{-8}$ as
compared with its terrestrial value and, hence, LPI is broken. In
view of the potentially important application of this discrepancy to
the fundamental physics, one has to be sure that the nonzero $\Delta \mu$ 
is not caused by some overlooked systematic errors. An obvious
way to tackle this problem is to use other molecular transitions
which have different sensitivity coefficients $Q_\mu$. It has
already been suggested to measure $\Lambda$-doublet lines of light
diatomic molecules OH and CH \citep{Koz09}, and microwave
inversion-rotational transitions in partly deuterated ammonia
NH$_2$D and ND$_2$H \citep{KLL10}.

In the present paper, we propose to use tunneling and rotation
transitions in the hydronium ion \hhho. Like ammonia, it also has a
double minimum vibrational potential. The inversion transitions
occurs when the oxygen atom tunnels through the plane of the
hydrogen atoms. This leads to an inversion splitting of the
rotational levels. The splitting of \hhho\ is very large,
$55.3462\pm0.0055$ cm$^{-1}$ \citep{LO85} 
as compared to $1.3$ cm$^{-1}$ splitting in \nhhh. Consequently, the ground-state
inversion-rotational spectrum of \hhho\ is observed in the
submillimeter-wave region \citep{PHL85,BDD85,VTM88}, whereas pure
inversion transitions in the far-infrared region
\citep{VVT89,YDPP09}.

\begin{figure}[t!]
\epsscale{1.1}
\plotone{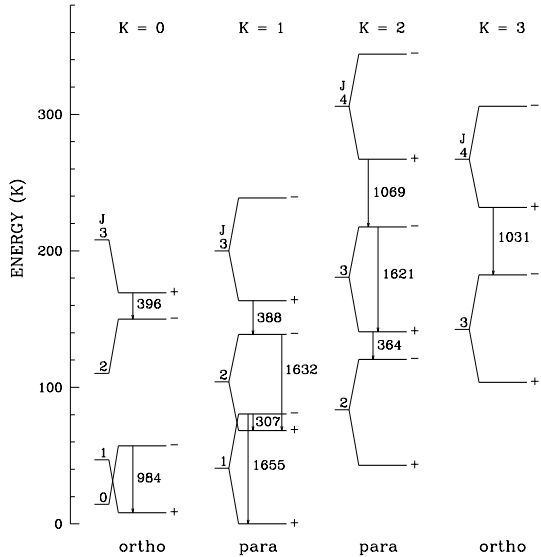}
\caption{The level scheme for \hhho. The depicted frequencies are in GHz. \label{fig1}}
\end{figure}

\hhho\ has both ortho- and para-modifications (see \fref{fig1}).
In the submillimeter,~-- the range accessible from high altitude
ground-based telescopes,~-- there are three low-lying transitions at
307, 364, and 388 GHz which belong to para-\hhho, and one
ortho-\hhho\ transition at 396 GHz. The 388 GHz line is, however,
blocked by water vapor in the atmosphere. The other lines were observed
in the interstellar molecular clouds
\citep{HCH86,WBB86,WMT91,vdT06,PvD92}.
The 364 GHz line was also
observed in two galaxies: \object{M~82} and \object{Arp~220} \citep{vdT08}. 
In far-IR, \hhho\ lines were detected from aboard the space
observatories at $\omega = 4.31$ THz \citep{TNP96}, 
$\omega = 1.66$, 2.97, and 2.98 THz \citep{GC01,LBS06,PBS07}, $\omega = 984$ GHz
\citep{GdL10}, and $\omega = 1.03, 1.07$, and 1.63 THz \citep{BBvD10}.

The observed transitions of \hhho\ arise in the warm ($T_{\rm kin}
\sim 100$ K) and dense ($n \approx 10^5-10^6$ cm$^{-3}$)
star-forming regions surrounding protostars where hydronium appears
to be one of the most abundant species with the abundance as high as
$X$(\hhho) $\approx 5\times10^{-9}$ \citep{WMT91,BBvD10}.

\section{Sensitivity coefficient of inversion transition}
\label{inv}

Sensitivity of the \hhho\ inversion transition to $\mu$-variation can be
estimated from the analytical Wentzel-Kramers-Brillouin (WKB)
approximation.
Following \cite{LL77}, we write for the inversion frequency 
(used units are $\hbar=|e|=m_e=1$):
\begin{equation}\label{inv1}
 \omega_\mathrm{inv}\approx \frac{2E_0}{\pi}\,\mathrm{e}^{-S},
\end{equation}
where $S$ is the action over classically forbidden region and $E_0$
is the ground state vibrational energy. Expression (\ref{inv1})
gives the following sensitivity to $\mu$-variation:
\begin{equation}\label{inv2}
 Q_\mathrm{inv}\approx \frac{S+1}{2}
 +\frac{S\,E_0}{2(U_\mathrm{max}-E_0)}\,,
\end{equation}
where $U_\mathrm{max}$ is the barrier hight and we are not using an
additional approximation $E_0=\omega_v/2$.

According to \cite{RNVH04} and \cite{DN06}, we can take
$U_\mathrm{max}=651$~\cm\ and $E_0\approx 400$~\cm. The inversion
frequency for \hhho\ is 55.3~\cm. Thus, Eqs. (\ref{inv1},\ref{inv2})
give:
\begin{equation}\label{inv3}
 S\approx 1.5\,,\qquad Q_\mathrm{inv}\approx 2.5\,.
\end{equation}

\begin{figure}[t!]
\epsscale{1.10}
\plotone{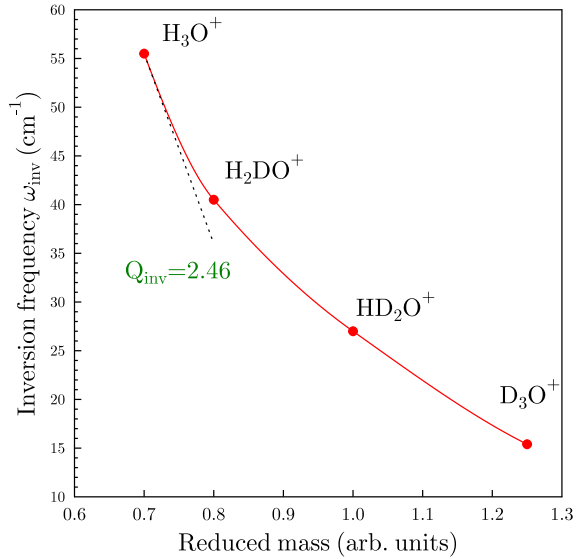}
\caption{Inversion frequency as a function of the reduced mass for
 hydronium ion isotopologues. \label{fig2}}
\end{figure}

\cite{DN06} report the inversion frequencies for \hhho,
H$_2$DO$^+$, HD$_2$O$^+$, and D$_3$O$^+$ to be 55.3 \cm, 40.5 \cm,
27.0 \cm, and 15.4 \cm, respectively. 
Neglecting the weak dependence of
the reduced mass, $m_r$, on the inversion coordinate, we take $m_r$ to be 0.7,
0.8, 1.0, and 1.25, respectively \citep{DN06}. 
Figure~\ref{fig2} shows the inversion frequency as a function of $m_r$.
From this plot we can estimate the sensitivity coefficient for \hhho\ to be:
\begin{equation}\label{inv4}
    Q_\mathrm{inv}\approx 2.46\,,
\end{equation}
which is in a perfect agreement with \Eref{inv3}. We can conclude that the inversion
transition in \hhho\ is almost two times less sensitive to $\mu$-variation,
than similar transition in \nhhh, where $Q_\mathrm{inv}=4.5$ \citep{FK07}.

\section{Sensitivity coefficients of mixed transitions}
\label{mix}

The spectrum of the rotational and inversion transitions of \hhho\ is
studied in \cite{YDPP09}. For the lowest vibrational state we can
write the simplified inversion-rotational Hamiltonian as:
\begin{eqnarray} \label{mix1}
 H\!\!\!\! & = &\!\!\!\! BJ(J+1) + (C-B)K^2 -D_J[J(J+1)]^2\quad  \nonumber \\
\!\!\!\! & &\!\!\!\! -D_{JK}J(J+1)K^2 -D_K K^4 + \dots \\ 
\!\!\!\! & &\!\!\!\! +\frac{s}{2} \left\{W_0 + W_J J(J+1) + W_K K^2 +\dots \right\}\, . \nonumber
\end{eqnarray}
Here we neglected higher terms of expansion in $J$ and $K$; $s=\pm 1$ for
symmetric and antisymmetric inversion state; total parity $p=(-1)^K s$.
Numerical values are given in \cite{YDPP09} (MHz):
\begin{equation}
\begin{array}{cccc}
  B  &  C-B  & D_J & D_{JK} \nonumber\\ 
 334406 & -148804 & 35 & -70 \nonumber\\  
 D_K &   W_0   & W_J & W_K    \nonumber\\
 41 & -1659350 & 5988 & -8458 \nonumber
\end{array}
\end{equation}
Note that we write Hamiltonian (\ref{mix1}) in such a way that terms which
determine inversion splitting are collected in the last line. Therefore, we
have the following relation with parameters used in \cite{YDPP09}:
\begin{eqnarray}\label{mix1a}
B &  = & \left[B(0^+)+B(0^-)\right]/2\,, \nonumber \\
W_J & = & B(0^+)-B(0^-)\,,
\end{eqnarray}
and similarly for $C-B$ and $W_K$. Parameters $D_J$, $D_{JK}$, and $D_K$ are
averaged over inversion states $s=\pm 1$.

To estimate sensitivities of the mixed transitions it is sufficient to
account for $\mu$-dependence of the dominant parameters $B$, $C$,
and $W_0$. It is clear that $B,\,C \sim\mu$ and $W_0$ scales as
$\mu^{Q_\mathrm{inv}}$. It follows, that for rotational part of the
energy we have $Q_\mathrm{rot}=1$ and for inversion part
$Q_\mathrm{inv}$ is given by \Eref{inv3} or (\ref{inv4}).
This leads to the expressions, used earlier for NH$_2$D
\citep{KLL10}:
\begin{equation}\label{mix2}
 \omega_\mathrm{mix} =\omega_\mathrm{rot}\pm\omega_\mathrm{inv}\, ,
\end{equation}
and
\begin{equation}\label{mix3}
 Q_\mathrm{mix}
 =\frac{\omega_\mathrm{rot}}{\omega_\mathrm{mix}}Q_\mathrm{rot}
 \pm\frac{\omega_\mathrm{inv}}{\omega_\mathrm{mix}}Q_\mathrm{inv}\, .
\end{equation}

We use Hamiltonian (\ref{mix1}) and expression (\ref{mix3}) to calculate the
frequencies and sensitivities of the mixed transitions. 
The obtained results are presented in \tref{tab_mix}.
Final results are very sensitive to the parameter $Q_\mathrm{inv}$.
A good agreement between two different estimates of $Q_\mathrm{inv}$
from \Eref{inv3} and from \fref{fig1} shows that this parameter is
known with 10\% accuracy, or better. In the next approximation, we
need to weight independently all terms of the Hamiltonian
(\ref{mix1}) with different scalings. However, this does not lead to
any significant changes in sensitivities $Q_\mu$ of the low-$J$
transitions from \tref{tab_mix}.

\begin{table*}
\begin{center}
\caption{Frequencies and sensitivities to $\mu$-variation of the
inversion-rotation transitions in \hhho. Experimental frequencies
are taken from \cite{JPL_Catalog,YDPP09}. \label{tab_mix}}
\begin{tabular}{cccccc r@{.}l c r c}
\tableline\tableline
\multicolumn{6}{c}{Transition} &\multicolumn{4}{c}{Frequency (MHz)} &\multicolumn{1}{c}{$Q_\mu$}\\
$J$&$K$&$s$&$J'$&$K'$&$s'$&\multicolumn{2}{c}{Exper.} 
& \multicolumn{1}{c}{error} &\multicolumn{1}{c}{\Eref{mix1}} \\
\tableline
  1 & 1  &$-1$&  2  &  1  &$+1$ &   307192&410& 0.05 &  307072&$+9.0$  \\
  3 & 2  &$+1$&  2  &  2  &$-1$ &   364797&427& 0.10 &  365046&$-5.7$  \\
  3 & 1  &$+1$&  2  &  1  &$-1$ &   388458&641& 0.08 &  389160&$-5.2$  \\
  3 & 0  &$+1$&  2  &  0  &$-1$ &   396272&412& 0.06 &  397198&$-5.1$  \\
  0 & 0  &$-1$&  1  &  0  &$+1$ &   984711&907& 0.30 &  984690&$+3.5$  \\
  4 & 3  &$+1$&  3  &  3  &$-1$ &  1031293&738& 0.30 & 1031664&$-1.4$  \\
  4 & 2  &$+1$&  3  &  2  &$-1$ &  1069826&632& 0.30 & 1071154&$-1.2$  \\
  3 & 2  &$-1$&  3  &  2  &$+1$ &  1621738&993& 2.00 & 1621326&$+2.5$  \\
  2 & 1  &$-1$&  2  &  1  &$+1$ &  1632090&98 &      & 1631880&$+2.5$  \\
  1 & 1  &$-1$&  1  &  1  &$+1$ &  1655833&910& 1.50 & 1655832&$+2.5$  \\
\tableline
\end{tabular}
\end{center}
\end{table*}

\section{Discussion and conclusions}
\label{dis}

We have shown above that the rest-frame frequencies of the
inversion-rotational transitions of \hhho\ are very sensitive to the
value of $\mu$. For a given transition from \tref{tab_mix},
$\omega_i$, with the sensitivity coefficient $Q_i$, the expected
frequency shift, $\Delta \omega_i/\omega_i$, due to a change in
$\mu$ is given by \citep{LML10}:
\begin{equation}\label{disEq1}
 \frac{\Delta \omega_i}{\omega_i}
 \equiv \frac{\tilde{\omega}_i - \omega_i}{\omega_i} =
 Q_i\frac{\Delta \mu}{\mu}\, ,
\end{equation}
where $\omega_i$ and $\tilde{\omega_i}$ are the frequencies
corresponding to the laboratory value of $\mu$ and to an altered
$\mu$ in a low-density environment, respectively.

By analogy with \Eref{eq1}, we can estimate the value of $\Delta \mu/\mu$ 
from two transitions with different sensitivity coefficients $Q_i$ and $Q_j$:
\begin{equation}\label{disEq2}
 \frac{\Delta \mu}{\mu} = \frac{V_j - V_i}{c(Q_i - Q_j)}\, ,
\end{equation}
where $V_j$ and $V_i$ are the apparent radial velocities of the
corresponding \hhho\ transitions.

Consider two lowest frequency transitions from \tref{tab_mix}: 
$1_1^- \rightarrow 2_1^+$ and $3_2^+ \rightarrow 2_2^-$ of
para-\hhho\ at, respectively, 307 and 364 GHz. Here $\Delta Q =
Q_{307} - Q_{364} = 14.7$, which is 4 times larger then the $\Delta Q$ 
value from the ammonia method. This means that the offset $\Delta V \sim 27$ \ms, 
detected in the ammonia method, should correspond to
the relative velocity shift between these transitions, $\Delta V =
V_{364} - V_{307}$, of about $100$ \ms.

Published results on interstellar \hhho\ allow us to put an upper
limit on $\Delta \mu/\mu$. The observations of the 307, 364, and 396
GHz lines carried out at the 10.4-m telescope of the Caltech
Submillimeter Observatory (CSO) by \cite{PvD92} and the observations of the 307 and 364
GHz lines at the 12-m APEX telescope (Atacama Pathfinder Experiment) by
\cite{vdT06} have accuracy of about 1 \kms\ that provides a limit
on $\Delta \mu/\mu < 2\times10^{-7}$, which is consistent with the
signal $\sim 3\times10^{-8}$  revealed by the ammonia method
\citep{LML10,LLH10}.

In order to check ammonia results we need to improve the accuracy of
\hhho\ observations by more than one order of magnitude. According to
\tref{tab_mix}, the uncertainties of the laboratory frequencies of
the transitions at 307, 364, and 396 GHz are, respectively, 50, 80,
and 45 \ms. Therefore, we also need a factor of few improvement of
the laboratory accuracy to be able to detect reliably an expected signal 
$\Delta V \sim 100$ \ms\ and to
check the non-zero ammonia results.

An important advantage of the hydronium method is that it is based
on only one molecule. In the ammonia method there is
unavoidable Doppler noise caused by relative velocity shifts due to
spatial segregation of \nhhh\ and other molecules. 
When using hydronium, the only
source of the Doppler noise may arise from possible kinetic
temperature fluctuations within the molecular cloud since the
submillimeter \hhho\ transitions have different upper level
energies: $E_u = 80, 139$, and 169 K for the 307, 364, and 396 GHz
transitions, respectively. It is also important that two \hhho\ transitions
have similar $Q$ values: $Q_{364} \approx Q_{396}$. This allows us to
control the Doppler noise and to measure accurately the relative position
of the 307 GHz line.

The analysis of other possible sytematic effects for hydronium is mostly
similar to what was done in detail for ammonia in Levshakov \etal\ (2010b).
The systematic shifts caused by pressure effects are about a few \ms, or lower.
As mentioned above in Sect.~\ref{sect-1}, the frequency shifts caused by external
electric and magnetic fields and by the cosmic black body
radiation-induced Stark effect are less or about 1 \ms\ for \nhhh. 

An additional source of
systematic for \hhho\ can come from the unresolved hyperfine structure (HFS)
in combination with possible non-thermal HFS populations in the ISM.
As noted by \cite{KR10}, HFS lines reduce the effective optical depth
of the molecular rotational transition by spreading the emission out over
a wider bandwidth.
To our knowledge, the HFS has not been resolved yet for H$_3{}^{16}$O$^+$ 
either in laboratory, or astronomical measurements.
An expected size of the hyperfine splittings can be estimated using analogy with ammonia. 
For the latter the main hyperfine
splitting is associated with the spin of nitrogen $\bm{I}_1$. 
The maximum hyperfine
splitting caused by the hydrogenic spin $\bm{I}$ is about 40 kHz
\citep{HT83}. This splitting includes interaction with the nitrogen spin
$\sim (\bm{I}_1\cdot\bm{I})$ and with molecular rotation 
$\sim (\bm{J}\cdot\bm{I})$. In hydronium, the oxygen nucleus is spinless and there is
only interaction with rotation $\sim (\bm{J}\cdot\bm{I})$. Hydronium
has similar electronic structure and close rotational constants to ammonia,
so its spin-rotational interaction should be $\la 40$ kHz.
Thus, we can expect less than 40 \ms\ of the hyperfine bandwidth for 300 GHz lines and
smaller splittings at higher frequencies. 
At $T_{\rm kin} \sim 100$~K,~-- a typical kinetic temperature
of warm and dense gas in the star-forming regions,~-- the
thermal width of the \hhho\ lines is comparable with this HFS splitting.

For para-\hhho, transitions at 307, 388 and 364 GHz have 3 HFS components each
(two transitions with $\Delta F = \Delta J$ are strong, and the remaining one is weak).
For ortho-\hhho, the 396 GHz transition has 9 HFS components
(4 strong, 3 weaker, and 2 weakest).

The difference in the magnitude of the energies of
the hyperfine and rotational transitions in molecular spectra in the 
sub-millimiter range is considerable: 
milli-Kelvin for the hyperfine levels and tens of Kelvin 
between rotational levels. 
This means that the HFS levels may be populated approximately in statistical equilibrium 
even if the rotational levels are not \citep{KR10}.
Therefore, we may suggest that the line centers of the sub-millimiter \hhho\
transitions are not affected significantly by relative populations of the hyperfine
levels and that the exepected velocity shifts are of a few \ms.

To sum up, the systematic shifts of the line centers 
caused by possible non-thermal HFS populations and pressure effects are likely
not larger than 10 \ms, which is about 10\% of the expected relative shift 
between the para-\hhho\ $J_K =
1_1^- \rightarrow 2_1^+$ and $3_2^+ \rightarrow 2_2^-$ transitions 
due to $\mu$-variation.  
A more accurate analysis of the HFS-induced systematics will be possible
after the HFS is either measured, or calculated theoretically.

Finally, we would like to note that other isotopologues of the hydronium ion
also must have large sensitivity coefficients $Q_\mu$. To use them
we need laboratory studies of the low-frequency mixed transitions in
the spectra of the partly deuterated hydronium ions H$_2$DO$^+$,
HD$_2$O$^+$, and in D$_3$O$^+$.

In the near future, high-precision measurements in the submillimeter and FIR ranges
with greatly improved sensitivity
will be available with the
Atacama Large Millimeter/submillimeter Array (ALMA), the Stratospheric Observatory For
Infrared Astronomy (SOFIA), the Cornell Caltech Atacama Telescope (CCAT), and others.
Thus, any further advances in exploring $\Delta \mu/\mu$ depend crucially on
accurate laboratory measurements ($\Delta \omega/\omega \la 10^{-8}$)
of relevant molecular transitions in the submillimeter and FIR ranges
where reliable spectroscopic data are still relatively poor.

\acknowledgments

We are grateful to V. Kokoouline for bringing inversion spectra of
\hhho\ to our attention. The work has been supported in part by the
RFBR grants 08-02-00460, 09-02-12223, and 09-02-00352, by the
Federal Agency for Science and Innovations grant NSh-3769.2010.2,
and by the Chinese Academy of Sciences,
grant No. 2009J2-6.

\end{document}